\documentclass[11pt]{article}
\usepackage[a4paper,margin=1in]{geometry}
\usepackage[T1]{fontenc}
\usepackage{graphicx}
\usepackage{bm}
\usepackage{mathtools}
\usepackage{amsmath,amssymb}
\usepackage{xcolor}
\usepackage{cite}

\newcommand{\abs}[1]{\left|#1\right|}
\newcommand{\Pec}{\mathrm{Pe}_{\mathrm{r}}}
\newcommand{\Teff}{T_{\mathrm{eff}}}
\newcommand{\letterlabel}[1]{\paragraph*{#1}}

\title{Arrhenius-Type Description and Noise-Composition Dependence\\
of Single-Vacancy Hopping in Active Brownian Crystals}

\author{Hiroto Tomida$^{*}$ and Yoshihiro Yamazaki\\
\small Department of Physics, Graduate School of Advanced Science and Engineering, Waseda University\\
\small $^{*}$E-mail: tomitomi.1610@akane.waseda.jp}
\date{}

\begin{document}
\maketitle

\begin{abstract}
We study vacancy-mediated hopping in a two-dimensional active Brownian-particle crystal with a single vacancy.
Under active driving, the hopping rates are not organized by the free-particle effective noise amplitude alone.
In contrast, in the passive limit, the hopping rate follows an Arrhenius-like activated trend with respect to the translational diffusion coefficient.
Effective-noise comparisons show systematic dependence on the active fraction, indicating that the composition of thermal and persistent active fluctuations affects the hopping kinetics.
These results demonstrate a breakdown of an effective-temperature description for microscopic defect-mediated transport in an active crystal.
\end{abstract}

% =========================
% Introduction
% =========================
% \section{Introduction}
\letterlabel{Introduction}: Active matter consists of units that continuously convert energy into motion and mechanical stresses, thereby sustaining dynamics far from thermal equilibrium \cite{Schweitzer2003,Ramaswamy2010,Marchetti2013,Bechinger2016}.  Dense active systems can acquire rigidity and long-lived cages, giving rise to solid-like active materials in both biological and synthetic settings \cite{Bi2016,Bechinger2016}.  An important question is how solid-state transport and relaxation are modified and characterized when the fluctuations are persistent and nonequilibrium rather than thermal \cite{Solon2015}.

Active Brownian particles (ABPs) provide a minimal particle-level model for this issue.  At high particle density, ABP systems can crystallize. Previous studies have constructed phase diagrams of dense active-particle
systems by classifying liquid, hexatic, and crystalline regimes and by tracking
how freezing and melting boundaries shift with activity
\cite{Bialke2012,Klamser2018}.  
% Other work has examined the melting process more
% microscopically, emphasizing the role of topological defects and two-step
% melting pathways in active Brownian crystals \cite{PaliwalDijkstra2020}.
Much of these studies focused on phase behavior and static order.  Here we instead treat a kinetic question in the solid-like regime: to what extent does an Arrhenius-like activated picture of vacancy-mediated hopping survive under nonequilibrium active driving?  Now we focus on the effective noise amplitude $\Teff$ motivated by the long-time diffusion of a free ABP.

In equilibrium crystals with some point defects, particle transport proceeds through point defects such as vacancies and interstitials, where localized rearrangements enable particles to hop between lattice-compatible sites \cite{Libal2007}.  Activity can reshape defect landscapes and rearrangement pathways \cite{vdMeer2016}.  It is so far not obvious whether persistent self-propulsion simply increases the noise amplitude or changes the hopping kinetics in a way that cannot be reduced to an effective-noise amplitude.

This question is closely related to barrier crossing by active particles. This is because vacancy hopping can be regarded as a barrier-crossing process in which a particle moves into the vacant site by overcoming the interaction-induced potential landscape formed by the surrounding solid-like particles. For an isolated active Brownian particle in two dimensions, with self-propulsion speed $v_0$, translational diffusion coefficient $D_t$,
and rotational diffusion coefficient $D_r$, the long-time diffusivity
contains an active contribution
$D_a=v_0^2/(2D_r)$, suggesting an effective-noise scale
$D_t+D_a$ \cite{Bechinger2016,tenHagen2011}. Active escape studies show, however, that persistent self-propulsion is not generally equivalent to thermal white noise at a higher temperature: active-Kramers or effective-equilibrium descriptions may involve activity-dependent effective potentials or apparent activation barriers, while finite persistence can alter escape pathways and invalidate a single-temperature mapping \cite{Geiseler2016,Sharma2017,Caprini2019,Militaru2021}.  
% Related active-solid experiments likewise show that active fluctuations can promote rearrangements across kinetic barriers and expose limits of a single effective-temperature description \cite{Ramananarivo2019,MassanaCid2024}.

We investigate how the known limitations of effective-noise descriptions appear in vacancy-mediated hopping in a controlled single-vacancy ABP crystal.
In reduced units where lengths are measured in units of the particle diameter
$\sigma$ and time in units of $D_r^{-1}$, and using the lattice-based activity
variable $\Pec=v_0/(aD_r)$ with $a$ measured in units of $\sigma$, the active
contribution to the free-particle diffusivity becomes
$D_a=a^2\Pec^2/2$.  We therefore introduce
\begin{equation}
\Teff \equiv D_t+\frac{a^2\Pec^2}{2},
\label{eq:teff_def_short}
\end{equation}
as a temperature-like comparison variable, but not as a thermodynamic
temperature of the interacting crystal.
Here $D_t$ denotes the reduced translational diffusion coefficient.

We find that the passive branch provides a useful activated reference, whereas
the active hopping rate is not controlled by the effective-noise amplitude
defined in Eq.~\eqref{eq:teff_def_short} alone.
Instead, the systematic deviations left by a $\Teff$-only description are empirically organized by the active fraction
\begin{equation}
  \chi \equiv \frac{a^2\Pec^2/2}{\Teff},
  \label{eq:chi_def_short}
\end{equation}
which measures the composition of the nominal effective noise.
Equivalently, the observed $\chi$-dependence may be viewed as a signature that activity modifies the apparent activation cost of a defect-mediated hop, rather than merely increasing the noise amplitude.
In this Letter, we use this controlled single-vacancy ABP crystal to demonstrate that vacancy-mediated hopping cannot be collapsed onto the single effective noise scale \(D_t + D_a\), indicating that the hopping dynamics also depends on the composition of the noise.
% =========================
% Model and hopping regime
% =========================
%\section{Model and defect-hopping regime}
\letterlabel{Model and Defect-Hopping Regime}: We numerically calculated the equations of motion for $N$ overdamped ABPs in two dimensions, which are given by
\begin{align}
\dot{\bm r}_i &= v_0\hat{\bm p}(\phi_i)+\mu\sum_{j\ne i}\bm F_{ij}(\bm r_i-\bm r_j)
+\sqrt{2D_t}\,\bm\xi_i(t),\label{eq:eom_r_letter}\\
\dot{\phi}_i &= \sqrt{2D_r}\,\eta_i(t),\label{eq:eom_phi_letter}
\end{align}
where \(\bm r_i\) and \(\phi_i\) denote the position and orientation angle of particle \(i\), respectively \(i=1,...,N\).
\(\hat{\bm p}(\phi_i)=(\cos\phi_i,\sin\phi_i)\) is the unit vector along its self-propulsion direction,
and \(\bm F_{ij}\) is obtained from the Weeks--Chandler--Andersen
(WCA) repulsive potential \cite{WCA1971},
\[
U_{\mathrm{WCA}}(r)=
\begin{cases}
4\epsilon
\left[
\left(\dfrac{\sigma}{r}\right)^{12}
-
\left(\dfrac{\sigma}{r}\right)^6
\right]
+\epsilon,
& r < 2^{1/6}\sigma, \\[6pt]
0,
& r \ge 2^{1/6}\sigma .
\end{cases}
\]
The corresponding force is given by
\[
\bm F_{ij}
=
-\nabla_i U_{\mathrm{WCA}}(r_{ij}),
\qquad
r_{ij}=|\bm r_i-\bm r_j|.
\]
Here, \(\sigma\) and \(\epsilon\) denote the length and energy scales of the WCA interaction, respectively; \(\sigma\) sets the particle-diameter, or excluded-volume, length scale.
We use $\sigma$ as the length unit and $D_r^{-1}$ as the time unit in our simulation, and set $\sigma=D_r=\mu=\epsilon=1$. Particles are initially placed on all \(N+1\) sites of a triangular lattice with lattice spacing \(a=1.15\sigma\). One particle is then removed to create a single vacancy.
The triangular lattice is replicated \(N_{\mathrm{rep}}\) times along each primitive
lattice direction. Our system has \(N_{\mathrm{rep}}=40\), and after
removing one particle to create a single vacancy, it contains
\(N=N_{\mathrm{rep}}^2-1=1599\) particles. The equations are integrated by the
Euler--Maruyama method with \(\Delta t=10^{-3}\).

\begin{figure*}[t]
  \centering
  \includegraphics[width=\linewidth]{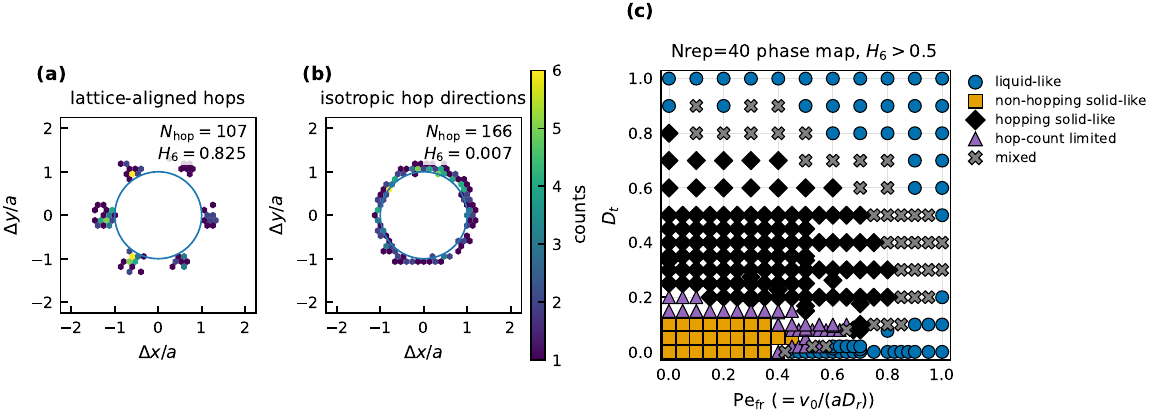}

  \caption{(Color online) Operational classification of the defect-hopping solid-like regime.
  \textbf{(a,b)} Representative hop-vector distributions for lattice-compatible hopping and disordered regimes.
  \textbf{(c)} State points are classified using the structural criterion $\overline{\psi}_6\ge0.80$ and the hopping criteria $H_6\ge0.50$ and at least 20 hops are detected.
  The hop-count cutoff is imposed to ensure stable directional statistics. Hop-count-limited points have too few detected hops to satisfy this cutoff and are therefore not used as hopping-solid-like data in the rate analysis. Mixed points indicate parameter sets for which independent runs give both solid-like and liquid-like/disordered classifications; these points are also excluded from the rate comparison.
  }
  \label{fig:regime_letter}
\end{figure*}

To compare rates only where hopping is lattice-compatible, we combine dynamical and structural diagnostics.  From unwrapped positions corrected by the instantaneous center of mass, we introduce the jump vector
\begin{equation}
  \bm J_i(t)=[\bm r_i(t)-\bm R_{\mathrm{CM}}(t)]-[\bm r_i^{\mathrm{ref}}-\bm R_{\mathrm{CM}}^{\mathrm{ref}}].
  \label{eq:jumpvec_letter}
\end{equation}
Here \(\bm R_{\mathrm{CM}}(t)=N^{-1}\sum_{j=1}^{N}\bm r_j(t)\) is the instantaneous
center-of-mass position computed from the unwrapped particle positions.
The reference quantities \(\bm r_i^{\mathrm{ref}}\) and
\(\bm R_{\mathrm{CM}}^{\mathrm{ref}}\) denote the particle and center-of-mass
positions at the corresponding reference time used for hop detection.
A hop candidate is detected when $\abs{\bm J_i}>d_{\mathrm{th}}$, with $d_{\mathrm{th}}=1.0\sigma$, and is confirmed only if the particle remains outside the threshold for five consecutive output frames.  The reference position is reset after confirmation.  We quantify lattice compatibility of the detected hop directions by
\begin{equation}
  H_6=\left|\left\langle e^{i6\varphi}\right\rangle_{\mathrm{hops}}\right|,
  \label{eq:h6_letter}
\end{equation}
where $\varphi$ is the polar angle of $\bm J$.  We also monitor local hexatic order,
\begin{equation}
  \psi_6(i)=\frac{1}{6}\sum_{j\in \mathrm{n.n.}(i)}e^{i6\theta_{ij}},\qquad
  \overline{\psi}_6=\left\langle |\psi_6(i)|\right\rangle_{i,t}.
  \label{eq:psi6_letter}
\end{equation}
A run is classified as hopping solid-like when $\overline{\psi}_6\ge0.80$, $H_6\ge0.50$, and at least 20 hops are detected.  Runs satisfying the structural criterion $\overline{\psi}_6\ge0.80$ but not the hopping criteria $H_6\ge0.50$ are classified as non-hopping solid-like, and runs with $\overline{\psi}_6<0.80$ are treated as liquid-like or disordered.  In this operational state map, mixed points denote parameter sets for which
independent runs yield both liquid-like/disordered and solid-like classifications.
They therefore indicate run-to-run variability rather than a distinct thermodynamic phase. Fig.~\ref{fig:regime_letter}(c) shows the hopping-solid-like regime identified by these criteria.

% =========================
% Rate analysis
% =========================
%\section{Arrhenius-type hop-rate test and effective noise amplitude}
\letterlabel{Arrhenius-Type Hop-Rate Test and Effective Noise Amplitude}: Let $N_{\mathrm{hop}}(t_0,t_0+t)$ be the total number of confirmed hops accumulated over all particles during a window of duration $t$.  We define the per-particle hop rate
\begin{equation}
  k=\lim_{t\to\infty}\frac{1}{N}\frac{\langle N_{\mathrm{hop}}(t_0,t_0+t)\rangle}{t}.
  \label{eq:rate_letter}
\end{equation}

In the passive limit $\Pec=0$, the dynamics reduces to an equilibrium overdamped Langevin system.  Since $D_t=\mu k_{\mathrm{B}}T$ in these units, vacancy hopping is expected to show an Arrhenius-like trend,
\begin{equation}
  k(D_t)\simeq A\exp\left(-\frac{\Delta E}{D_t}\right).
  \label{eq:arrhenius_letter}
\end{equation}

\begin{figure*}[t]
  \centering
  \includegraphics[width=\linewidth]{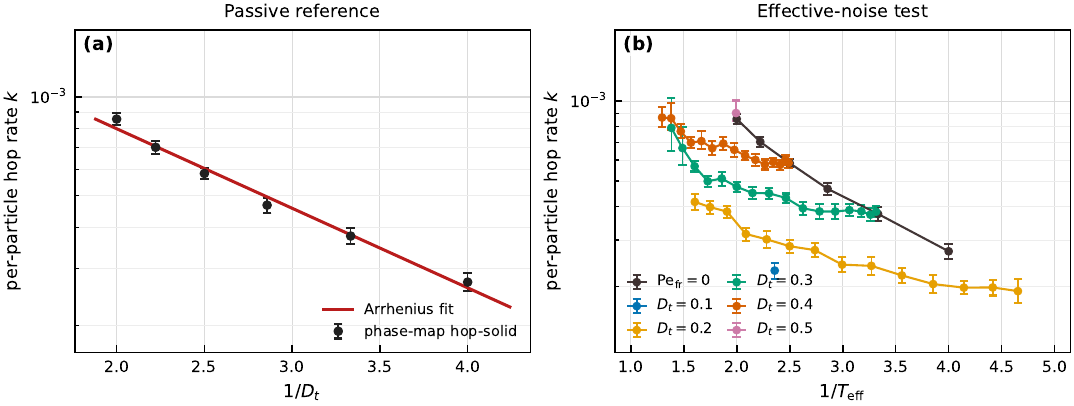}
  \caption{(Color online) Hop-rate tests within the retained hopping-solid-like regime.  \textbf{(a)} Passive rates follow an Arrhenius-like reference $k=A\exp(-\Delta E/D_t)$, with $\Delta E=0.521$ and $A=0.0288$.  \textbf{(b)} Effective-noise test: even using $\Teff=D_t+a^2\Pec^2/2$, active and passive data do not fall on a single curve versus $1/\Teff$.  Error bars denote two-sided 95\% percentile bootstrap confidence intervals over independent runs.}
  \label{fig:rates_letter}
\end{figure*}

For the retained passive hopping-solid-like state points, the run-averaged rates follow this activated reference, with $\Delta E=0.521$, $A=0.0288$, and weighted $R^2=0.985$, see Fig.~\ref{fig:rates_letter}(a).  This passive branch provides the baseline against which active hopping is compared.

For a free ABP, the long-time diffusivity suggests the effective noise amplitude in Eq.~\eqref{eq:teff_def_short}.  This quantity is a useful comparison variable, but it is not a thermodynamic temperature of the interacting crystal; it does not encode persistence time, directionality, or nonequilibrium path structure.  If hopping were fully controlled by $\Teff$ given by Eq.~\eqref{eq:teff_def_short}, all retained rate data should fall on a single curve when plotted versus $1/\Teff$.  They do not.  As shown in Fig.~\ref{fig:rates_letter}(b), active sweeps at fixed $D_t$ remain systematically separated from each other and from the passive branch, even at comparable $\Teff$.

\begin{figure}[t]
  \centering
  \includegraphics[width=\linewidth]{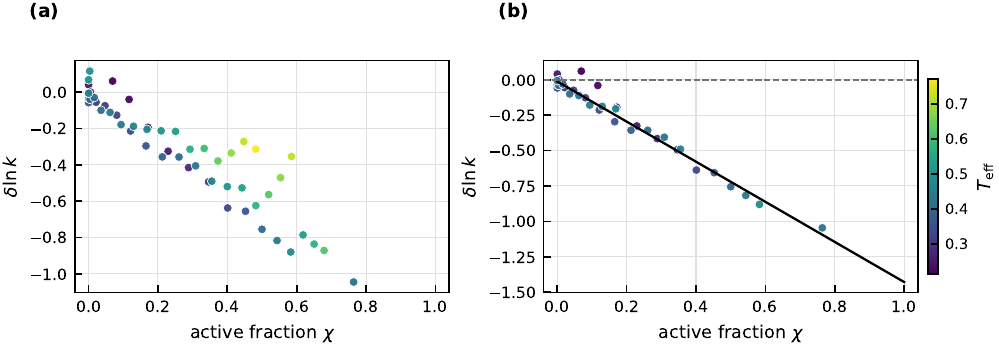}
    \caption{(Color online)
    Noise-composition dependence beyond a scalar effective-noise description.
    The passive-normalized residual
    $\delta\ln k$
    is plotted against the active fraction
    $\chi=(a^2\Pec^2/2)/\Teff$.
    Color indicates $\Teff$.
    Panel \textbf{(a)} shows all retained hopping-solid-like state points, whereas
    panel \textbf{(b)} focuses on the low-effective-noise subset $\Teff<0.5$.
    The solid line in panel \textbf{(b)} is a weighted linear fit to this subset,
    and the dashed line denotes zero residual.
    }

  \label{fig:chi_letter}
\end{figure}

At fixed \(\Teff\), changing the active fraction \(\chi\) changes the mean hop rate,
showing that the fluctuation composition cannot be eliminated in favor of an
effective-noise amplitude. We quantify the deviation from a \(\Teff\) description by subtracting the passive Arrhenius prediction evaluated at the same \(\Teff\),
\begin{equation}
  \delta\ln k=\ln k-\ln k_{\mathrm{passive}}(\Teff).
  \label{eq:residual_letter}
\end{equation}
A one-parameter $\Teff$ description would predict no systematic
dependence of the residual $\delta\ln k$ on $\chi$.  Instead, the
retained hopping-solid-like data show a clear residual dependence on
$\chi$, as shown in Fig.~\ref{fig:chi_letter}(a).  When restricted to
the low-$\Teff$ subset in Fig.~\ref{fig:chi_letter}(b), adding $\chi$
to the weighted log-rate model raises the weighted $R^2$ to $0.990$.
This $\chi$-augmented form therefore provides a minimal
phenomenological description of the systematic deviations that remain
after the $\Teff$ dependence is removed.

% =========================
% Discussion and conclusion
% =========================
%\section{Discussion and conclusion}
\letterlabel{Discussion and Conclusion}: The passive activated picture remains a useful reference for the retained hopping-solid-like branch.  When persistent active forcing is included, however, the hop rate is not uniquely organized by the free-particle effective noise amplitude $\Teff$ alone.  $\Teff$ captures the broad tendency that hopping is suppressed as the total noise amplitude is reduced, but it is insufficient as a quantitative one-parameter description.  

The systematic deviations left by $\Teff$ alone can be organized as noise-composition dependence characterized by the active fraction $\chi$. 
We note that all data used in this analysis are restricted to the hopping-solid-like state. This is necessary because the large-$\chi$ region also includes liquid-like states, as seen in Fig.~1(c). Accordingly, the lack of data points for $\chi>0.8$ in Fig.~3(b) simply indicates that those parameter sets are outside the hopping-solid-like state and fall into the liquid-like region.

Importantly, the effective-temperature viewpoint remains relevant. In the low-effective-noise regime, $\Teff<0.5$, where vacancy-mediated hops remain rare and lattice-compatible, the remaining deviations from a $\Teff$-only description are well captured by the active fraction $\chi$. Thus, the hopping rate can be viewed as being controlled by a leading Arrhenius-like scale set by $\Teff$, together with a correction associated with the composition of the fluctuations. A simple interpretation is obtained by writing the rate in an activated form with an activity-dependent apparent activation cost,
\begin{align}
  k&\simeq A\exp\!\left[-\frac{\Delta E_{\mathrm{app}}(D_a)}{\Teff}\right],
  \label{eq:activated_app_rate}\\
  \Delta E_{\mathrm{app}}(D_a)&\simeq \Delta E_0+\lambda D_a,
  \qquad D_a=\frac{a^2\Pec^2}{2}.
  \label{eq:active_barrier_cost}
\end{align}
Substituting Eq.~\eqref{eq:active_barrier_cost} into
Eq.~\eqref{eq:activated_app_rate}, and using \(\chi=D_a/\Teff\), gives $\ln k=\ln A-\Delta E_0/\Teff-\lambda\chi$.  Thus the failure of a \(\Teff\) collapse can be understood as the competition between the following two effects; activity increases the effective noise amplitude, but it also increases the apparent activation cost for lattice-compatible vacancy hopping.
It should also be emphasized that Eqs. (11) and (12) provide an effective description over much of the hopping-solid-like regime, although they are not expected to account for every parameter region. As shown in Fig. 3(a), there remain regions where the rate cannot yet be quantitatively organized by $\Teff$ and $\chi$ alone.
This result should be interpreted as a controlled extension of an effective-temperature description. The present result shows that, for microscopic activated transitions in an active crystal, this role is only partly captured by $\Teff$: a quantitative description also requires the composition variable $\chi$, which encodes how the fluctuations are generated. 
At the current stage, this conclusion is deliberately restricted to a controlled single-vacancy, fixed-system-size setting, which isolates one microscopic transport mechanism. Future work should test system-size, vacancy-concentration, and persistence-time dependence. The present results show that even this elementary defect process is not described by fluctuation amplitude alone, but can be organized by an extended description involving both $\Teff$ and $\chi$.

%\section{Acknowledgment}
{\bf Acknowledgment}\quad The authors are grateful to Mr. K. Okui and Dr. K. Taga for useful comments and encouragement.

% =========================
% References
% =========================


\begingroup
\footnotesize
\begin{thebibliography}{10}

\bibitem{Schweitzer2003}
F.~Schweitzer: {\em Brownian Agents and Active Particles: Collective Dynamics
  in the Natural and Social Sciences} (Springer Series in Synergetics. Springer
  Berlin, Heidelberg, 2003), Springer Series in Synergetics.

\bibitem{Ramaswamy2010}
S.~Ramaswamy: Annual Review of Condensed Matter Physics {\bfseries 1} (2010)
  323.

\bibitem{Marchetti2013}
M.~C. Marchetti, J.-F. Joanny, S.~Ramaswamy, T.~B. Liverpool, J.~Prost, M.~Rao,
  and R.~A. Simha: Reviews of Modern Physics {\bfseries 85} (2013) 1143.

\bibitem{Bechinger2016}
C.~Bechinger, R.~Di~Leonardo, H.~L{\"o}wen, C.~Reichhardt, G.~Volpe, and
  G.~Volpe: Reviews of Modern Physics {\bfseries 88} (2016) 045006.

\bibitem{Bi2016}
D.~Bi, X.~Yang, M.~C. Marchetti, and M.~L. Manning: Physical Review X
  {\bfseries 6} (2016) 021011.

\bibitem{Solon2015}
A.~P. Solon, Y.~Fily, A.~Baskaran, M.~E. Cates, Y.~Kafri, M.~Kardar, and
  J.~Tailleur: Nature Physics {\bfseries 11} (2015) 673.

\bibitem{Bialke2012}
J.~Bialk{\'e}, T.~Speck, and H.~L{\"o}wen: Physical Review Letters {\bfseries
  108} (2012) 168301.

\bibitem{Klamser2018}
J.~U. Klamser, S.~C. Kapfer, and W.~Krauth: Nature Communications {\bfseries 9}
  (2018) 5045.

\bibitem{Libal2007}
A.~Lib{\'a}l, C.~Reichhardt, and C.~J. Olson~Reichhardt: Physical Review E
  {\bfseries 75} (2007) 011403.

\bibitem{vdMeer2016}
B.~van~der Meer, L.~Filion, and M.~Dijkstra: Soft Matter {\bfseries 12} (2016)
  3406.

\bibitem{tenHagen2011}
B.~ten Hagen, S.~van Teeffelen, and H.~L{\"o}wen: Journal of Physics: Condensed
  Matter {\bfseries 23} (2011) 194119.

\bibitem{Geiseler2016}
A.~Geiseler, P.~H{\"a}nggi, and G.~Schmid: The European Physical Journal B
  {\bfseries 89} (2016) 175.

\bibitem{Sharma2017}
A.~Sharma, R.~Wittmann, and J.~M. Brader: Physical Review E {\bfseries 95}
  (2017) 012115.

\bibitem{Caprini2019}
L.~Caprini, U.~Marini Bettolo~Marconi, A.~Puglisi, and A.~Vulpiani: The Journal
  of Chemical Physics {\bfseries 150} (2019) 024902.

\bibitem{Militaru2021}
A.~Militaru, M.~Innerbichler, M.~Frimmer, F.~Tebbenjohanns, L.~Novotny, and
  C.~Dellago: Nature Communications {\bfseries 12} (2021) 2446.

\bibitem{WCA1971}
J.~D. Weeks, D.~Chandler, and H.~C. Andersen: The Journal of Chemical Physics
  {\bfseries 54} (1971) 5237.

\end{thebibliography}

\endgroup
\end{document}